\documentclass[prd,  twocolumn, 10pt]{revtex4-1}

\usepackage{amsmath}
\usepackage{amsfonts}
\usepackage{amssymb}
\usepackage{graphicx}
\usepackage{epsfig}
\usepackage{mathrsfs}
\usepackage{bm}
\usepackage{microtype}
\usepackage{hyperref}
\usepackage{mathrsfs}

\usepackage{sirimacro}

\newcommand\HH{\mathscr{H}}
\renewcommand\eps\epsilon
\def\tdN{\tilde{N}}

\DeclareSymbolFont{extraup}{U}{zavm}{m}{n}
\DeclareMathSymbol{\varheart}{\mathalpha}{extraup}{86}

\begin{document}

\title{Inflationary e-folding and the implications for gravitational-wave detection}

\author{Siri Chongchitnan}

\date{\today}
\affiliation{E. A. Milne Centre for Astrophysics, University of Hull,  HU6 7RX, United Kingdom.}

\email{s.chongchitnan@hull.ac.uk}

\begin{abstract}
We demonstrate that the approximation for the number of inflationary e-folds commonly used in the literature can lead to highly inaccurate predictions for the amplitude of primordial gravitational waves. We show that such an approximation can lead to perfectly viable inflation models being falsely ruled out by direct or indirect gravitational-wave measurements. We illustrate this point using a new class of inflation models which include the power-law potential, $V(\phi)\sim\phi^k$, as the simplest limit. 
These models are simple to construct without using the slow-roll approximation, and are consistent with $2\sigma$ constraints from Planck. Crucially, these models may suffer from an order-of-magnitude error in the prediction for the gravitational-wave amplitude if the common definition of e-folding is used. Our findings have strong implications for the classes of inflation models that can be ruled out by future space-based laser interferometers such as BBO and DECIGO.

\end{abstract}

\maketitle
\section{Introduction}

Cosmological inflation refers to the period of accelerated expansion in the early Universe  \cite{guth, linde, starobinsky2}. This simple idea has profound implications for our understanding of the Universe. Remarkably, in one fell swoop, it solves several longstanding cosmological conundrums, namely the horizon, flatness and monopole problems. The simplicity of constructing a working inflation model (at least in the phenomenological sense) means that inflation remains, even after almost four decades, the leading theory that can explain the origin of large-scale structures of the Universe, right down to the statistics of small-scale anisotropies in the cosmic microwave background (CMB). See \cite{LLbook,baumann, martin} for comprehensive reviews of inflation.

As inflation progresses into the next decade, new ambitious  experiments will endeavour to  uncover its  `smoking-gun' signatures that have hitherto eluded detection. The most tantalising of these signatures is the stochastic background of primordial gravitational waves (GW), which can be measured either directly using laser interferometers or indirectly via the measurement of B-mode polarization in the CMB. 

The predicted amplitude of this primordial GW background depends on \ii{when} the Fourier modes corresponding to the GW frequencies was expelled from the Hubble radius during inflation \footnote{The amplitude of superhorizon perturbations is frozen at least in the standard models of inflation presented in this paper. See \cite{leach} for interesting case studies.}. Modes that exited the Hubble radius at different time during inflation are associated with GW at different frequencies today. This so-called `horizon-exit' time can be equivalently parametrized using the \ii{e-fold} number, defined as the ratio of the Hubble radius $(aH)^{-1}$ at the exit time, $t$, and at the end of inflation, \ie
\ba\td{N}(t)\equiv \ln{a(t\sub{end})H(t\sub{end})\over a(t)H(t)}.\lab{two}\ea
For example, to solve the horizon problem, we require the Fourier modes corresponding to the CMB scales ($k\approx0.002$ Mpc$^{-1}$) to exit the horizon roughly $60$ e-folds before inflation ends.

Nevertheless, it has become common practice to approximate the number of e-folds by stipulating that the Hubble parameter, $H$, is essentially constant during inflation (in comparison with the scale factor $a$), and consider instead the logarithmic change in the scale factor alone:
\ba N(t)\equiv \ln{a(t\sub{end})\over a(t)}.\lab{badN}\ea
This approximation is sufficiently accurate for many inflation models in the literature, particularly for those models in which the `slow-roll' approximation holds. However, since Eq. \ref{badN} does not contain any reference to the `horizon' (\iee the Hubble radius), it follows that requiring ``$N=60$" will not necessary solve the horizon problem.

In \cite{meinf}, we constructed a simple inflation model in which the $N$ and $\tdN$ framework could differ enough so that the predictions for $r$ corresponding to either $N=60$ or $\tdN=60$ are inconsistent at around $5\%$ level. We cautioned that even such a small discrepancy could have significant implications for what we will learn from future GW experiments such as LISA, BBO and DECIGO \cite{lisa,bbo,decigo}.

In this work, we present an extension of the construction in \cite{meinf} and demonstrate that the discrepancy in the predicted GW amplitude, by assuming that $N=60$ or $\tdN=60$, could in principle be \ii{arbitrarily large}. These models are straightforward to construct and are consistent at $2\sigma$ level with the observational constraints from Planck \cite{planck15}. We argue that such a discrepancy should not be treated as an pathological anomaly, but rather a stark reminder that using the approximate definition, $N$, for the number of e-folds carries with it potentially dangerous yet completely avoidable inaccuracies.

For the rest of this work, where necessary we will refer to $\tdN$ as the \ii{physical} e-fold, and $N$ as the \ii{approximate} e-fold. We will work with the reduced Planck unit in which $ m\sub{Pl}/\sqrt{8\pi }=1$. We will only consider single-field inflation.


\section{A new framework of inflation model building}

In \cite{meinf}, we presented an alternative framework for inflation model building in which the physical number of e-folds, $\tdN$, is the key temporal parameter (as opposed to the approximation $N$). We summarise the main results below.

Our formalism begins with modelling of the inverse Hubble radius, $\HH$, defined as 
\ba \HH(\phi) \equiv aH,\ea
where $\phi$ is the inflaton value in unit of the Planck mass. We shall see below that $\HH(\phi)$ completely determines the inflaton potential $V(\phi)$. It is useful to define the series of variables, $E_n$, where
\ba E_n (\phi) &\equiv {\HH^{(n)}(\phi)\over \HH(\phi)}.\ea
The first few values of $E_n$ $(n=1,2,3)$  will determine the next-to-leading-order approximation for the inflationary observables $r$ (the tensor-to-scalar ratio) and $n_s$ (the spectral index of scalar perturbations). 

By definition, inflation occurs as long as the Hubble radius shrinks, \ie
\ba\text{Inflation}\iff{\D\over \D t}\HH>0 \iff E_1>0.\ea
Therefore, in our framework, an inflation model can be constructed using an \ii{increasing} function $\HH(\phi)$, with inflation ending at a maximum point. This is an analogous (but simpler) construction to the Hamilton-Jacobi formalism \cite{liddle+,lidseybig} in which a decreasing function $H(\phi)$ is the key variable, together with the `Hubble slow-roll' parameters
\ba \epsilon &\equiv 2\left({H'\over H}\right)^2,\quad \eta \equiv2{H''\over H},\quad \xi\equiv4{ H^\prime H^{\prime\prime\prime}\over
H^2}\ff,\nn\\
^\ell\lambda_H\ &\equiv 2^\ell{(H')^{\ell-1}\over H^\ell} H^{(\ell+1)}(\phi)\ff\label{flowparam}. \ea
In the Hamilton-Jacobi formalism,
\ba\text{Inflation}\iff{\D\over \D t} H<0 \iff \eps<1.\ea
The $\HH$ parameter is related $H$ and the potential $V(\phi)$ by the following flowchart:
\ba \HH(\phi) \quad \Longrightarrow\quad E_1=&\HH^\pr/\HH\lab{flowchart}\\
&\big\Downarrow\nn\\
 \eps  =& {2\over\bkt{E_1+\sqrt{(E_1)^2+2}}^2}\notag\\
&\big\Downarrow\nn\\
 H(\phi)=H\sub{end}&\exp\bkt{-\int_{\phi\sub{end}}^\phi\sqrt{\eps/2}\D\phi}\notag \\
 &\big\Downarrow\nn\\
  V(\phi)=&H^2(3-\eps)\nn
\ea

The variable $\HH$ is connected to the physical e-fold, $\tdN$, simply by
\ba\td N(\phi)&=\ln\bkt{\HH(\phi\sub{end})\over\HH(\phi))},\nn\\
\diff{\tilde N}{\phi}&=-{\mathscr{H}^\pr\over\mathscr{H}}=-E_1.\lab{solve-ana}\ea
In contrast, the approximate e-fold, $N$, is a more natural temporal variable in the Hamilton-Jacobi formalism, since we have
\ba \diff{N}{\phi}={H\over2H^\pr}=-{1\over\sqrt{2\eps}}.\lab{solve-n}\ea
This illustrates another advantage of the $\HH$ formalism: $\tdN(\phi)$ can be evaluated exactly (often by hand) via Eq. \ref{solve-ana}, whereas Eq. \ref{solve-n}  does not usually have an analytic solution.

Finally, the observables $r$ and $n_s$ can be evaluated using the usual next-to-leading expressions in terms of the Hubble-slow-roll parameters \cite{lidseybig}:
\ba r &\simeq 16\epsilon[1-C(\sigma+2\epsilon)]\ff, \label{r2}\\ 
n_s&\simeq1+\sigma-(5-3C)\epsilon^2-{1\over4}(3-5C)\sigma\epsilon  +{1\over2}(3-C)\xi\nn,
\ea 
where  $C=4(\ln2+\gamma)-5\simeq0.0814514$ (with $\gamma$ the Euler-Mascheroni constant). The Hubble-slow-roll parameters are related to $E_n$ by:
\ba \eps  &= {2\over\bkt{E_1+\sqrt{(E_1)^2+2}}^2},\lab{hsr}\\
\eta &= {\eps(2E_2 + 3) -1\over 1+\eps},\nn\\
\xi &= {\eps\over (1+\eps)^3}\bigg(3\eps^3-2\sqrt2\eps^{5/2}E_3-2E_2\eps^2+8\eps (E_2)^2-3\eps^2 \ldots\notag\\
&-4\sqrt2E_3\eps^{3/2}+28\eps E_2+17\eps-2\sqrt{2}E_3\sqrt{\eps}-2E_2 -9 \bigg).\nn
\ea



\section{The generalised Gaussian model}

We now present a detailed study of an inflation model parametrized by
\ba\HH(\phi)\propto e^{-(\alpha\phi)^n}\lab{gen},\ea
where the constant $\alpha>0$. We shall refer to Eq. \ref{gen} as the \ii{generalised Gaussian} model.

In this model, $n$ is an even positive integer, so that $\HH(\phi)$ is an increasing function along the branch $\phi<0$ (which is not problematic thanks to the even symmetry of $\HH$ and the $t\to-t$ transformation). It is possible to extend the range of $n$ to any positive real number by performing the symmetrization $\HH\propto  \exp\bkt{-\bkt{\alpha|\phi|}^n}$. Inflation ends at the maximum point $\phi=0$.

The Gaussian case $n=2$ is particularly interesting because, as shown in \cite{meinf}, it gives almost exactly the same predictions in the $n_s$-$r$ plane as those from the well-known power-law (monomial) models 
\ba V(\phi)\propto \phi^k, \quad k = 1/2\alpha^2.\lab{PL}\ea
In fact, we will show later that this is no coincidence: the inflaton potential for $n=2$ does in fact reduce to the power-law form to a good approximation. 

The coefficients $E_i$ can be expressed in terms of $\tdN$ as:
\ba 
E_1 &= n \alpha  \td N^{1-{1\over n}},\\
E_2 &=n\alpha^2\td N^{1-{2\over n}}\bkt{n\td N-n+1},\\
E_3 &= n\alpha^3 \td N^{1-{3\over n}}\bkt{n^2\td N^2-3n(n-1)\td N+(n-1)(n-2)}.
\ea

When these expressions are evaluated at $\tdN=\tdN_*$ (say, $\tdN=60)$, the corresponding inflaton field value $\phi_*$ is given by
\ba \phi_*=-\alpha^{-1}\tdN_*^{1/n}.\ea
For the rest of this work, we will assume that CMB-scale perturbations were generated when $\tdN_*=60$, although our main results are insensitive to this choice of $\tdN_*$.

\subsection{$r$ and $n_s$}

\begin{figure} 
   \centering
   \includegraphics[width=3.5in]{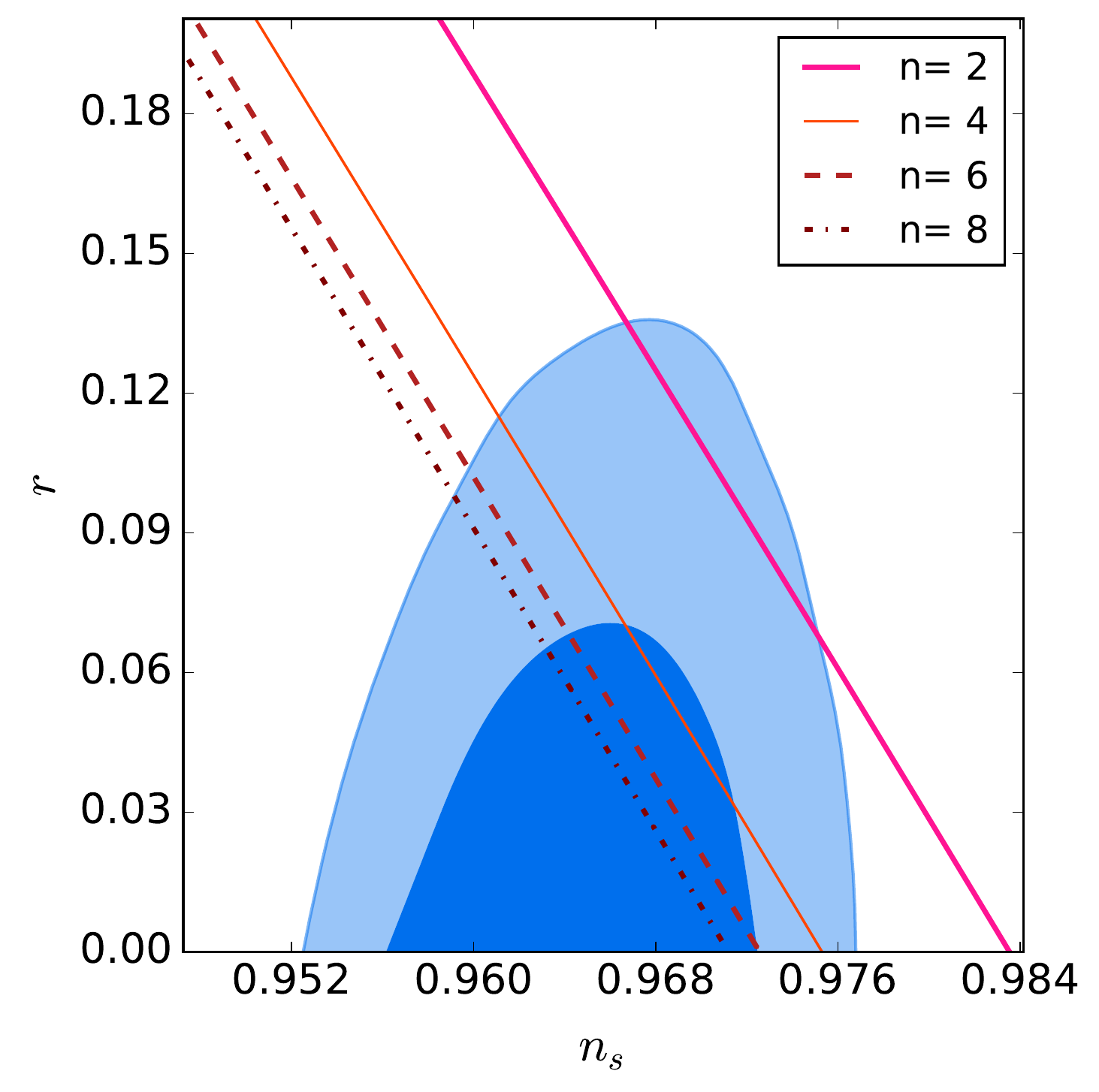} 
   \caption{The predictions for $n_s$ (the scalar spectral index) against  $r$ (the tensor-to-scalar ratio) for the generalised Gaussian model $\HH\propto\exp(-(\alpha\phi)^n)$ with $n=2,4,6,8$, assuming $\tdN_*=60.$ The values of $\alpha$ consistent with the $2\sigma$ constraints from Planck (blue contours) are given in the Table I.}
   \label{fig_varyn}
\end{figure}

Figure \ref{fig_varyn} shows the prediction in the $n_s$-$r$ plane for the generalised Gaussian model with $n=2,4,6$ and 8 for a wide range of $\alpha$. The contours show $1$ and $2\sigma$ constraints from Planck (temperature anisotropies + low $\ell$ polarization). The values of $\alpha$ that are consistent with these constraints are shown in Table \ref{tabalpha}.

\begin{table}[h]
\begin{tabular}{|c|c|}
\hline
$\quad n\quad$& $\alpha$\\
\hline
2 & $\quad0.5<\alpha<0.7\quad$\\
4 & $\alpha>0.096$\\ 
6 & $\alpha>0.048$\\
8 & $\alpha>0.031$\\
\hline
\end{tabular}
\caption{The values of the parameters $n$ and $\alpha$ for which the generalised Gaussian model $\HH\propto e^{-(\alpha\phi)^n}$ is consistent with Planck's $2\sigma$ constraints in the $n_s$-$r$ plane.}
\lab{tabalpha}
\end{table}
It can be shown that the lines in Fig. \ref{fig_varyn} can be described by the equations
\ba 
r&\approx {16\over 2(\alpha n\tdN^{1-1/n})^2+1},\\
n_s&\approx A + Br,
\ea
where the constants $A$ and $B$ depend on $\alpha$ and $n$. For $n=2$, $B\approx-1/8$ as is well known for power-law potentials. Despite the appearance, the four lines in Fig. \ref{fig_varyn} are not parallel (the horizontal axis is far more highly resolved).

\subsection{Is it safe to assume $\tdN=N$?}

We now demonstrate the approximate e-fold number, $N$, can be a highly inaccurate measure of inflation.

The relationship between $N$ and $\tdN$ is
\ba N={\tdN\over2} +{1\over2}\int_0^{\tdN} \sqrt{1+{2\over E_1^2}}\D\tdN.\lab{NvsN}\ea

For models in which the slow-roll approximation holds, $\eps\ll1$ throughout the duration of inflation, and from \re{hsr} this means $E_1\gg1$. Thus, \re{NvsN} suggests that $N\approx\tdN$ in the slow-roll limit. Let us see if this is the case for the generalised Gaussian model. 

\begin{figure} 
   \centering
   \includegraphics[width=3.3in]{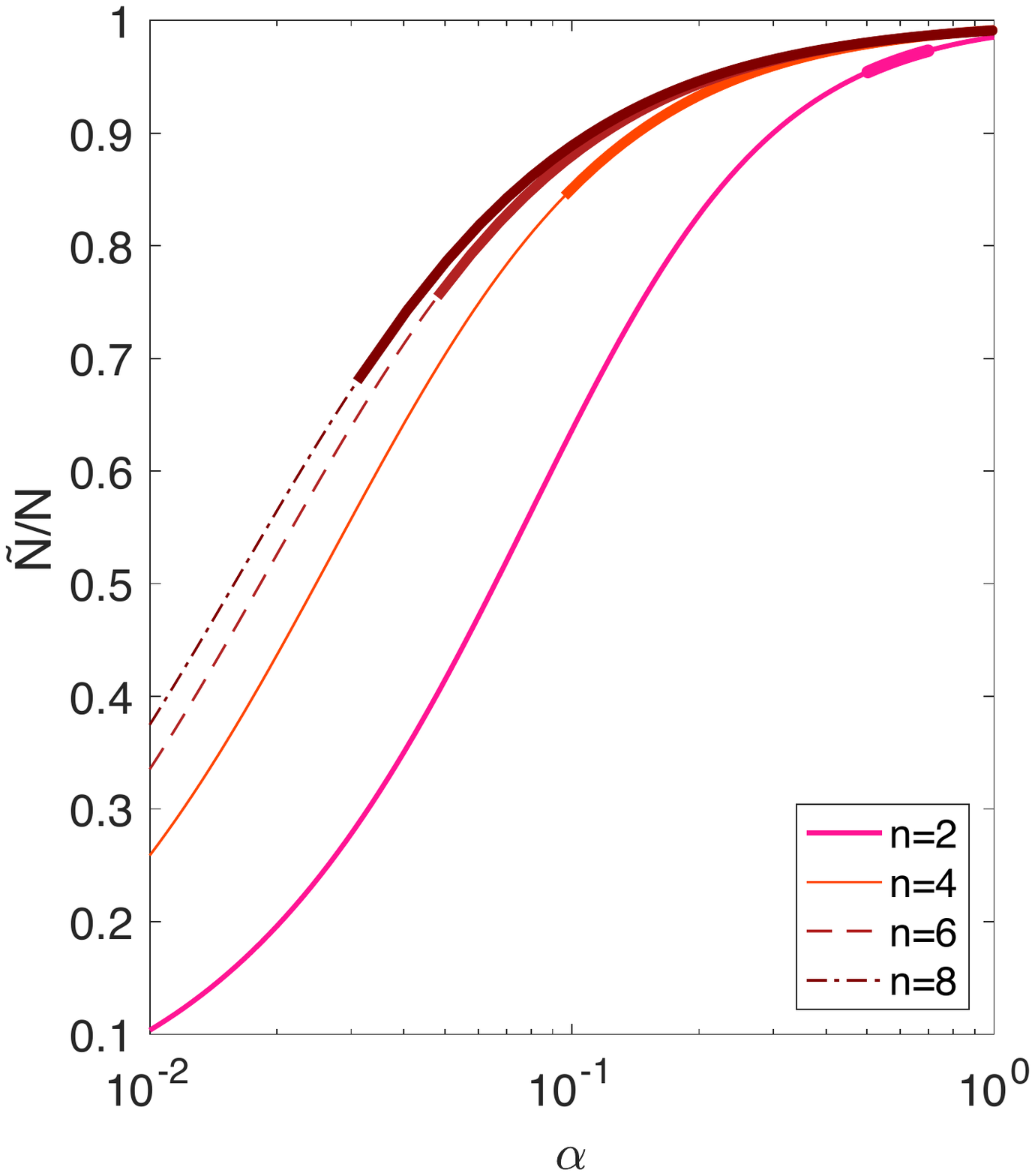} 
   \caption{The ratio between the physical and the approximate e-folds for the model $\HH=\exp(-(a\phi)^n)$, for $n=2,4,6,8$, assuming $\tdN_*=60.$ The thick solid portion of each line corresponds to the range of values of $\alpha$ which are consistent with the $2\sigma$ constraints from Planck. }
   \label{fig_thickthin}
\end{figure}

Fig. \ref{fig_thickthin} shows a plot of the ratio $\tdN/ N$ for $n= 2, 4, 6$ and $8$ as a function of the parameter $\alpha$, assuming $\tdN_*=60$. On each curve, the thick solid portion  corresponds to the range of values of $\alpha$ which are consistent with the $2\sigma$ constraints from Planck (see Table \ref{tabalpha}). We can see, for example, that for the case $n=2$, the approximation $\tdN\approx N$ is fairly accurate for Planck-consistent values of $\alpha$, with error of just a few percent (although interestingly, the equality $\tdN=N$ is ruled out at $2\sigma$). This is consistent with conventional wisdom from introductory cosmology that the slow-roll approximation can be safely applied to power-law potentials.


Unfortunately, the same error grows to an unacceptable level for larger values of $n$. With $n=8$, observation allows $\tdN$ and $N$ to differ by over $30\%$, signifying a strong breakdown in the slow-roll approximation.

It is interesting to note that the relationship \re{NvsN} can be integrated directly in the power-law case $n=2$ \cite{meinf}. For other values of $n$, the integral can be expressed in terms of the hypergeometric function.

We now consider how strongly this discrepancy between $\tdN$ and $N$ translates into errors in the $n_s$-$r$ plane. 

\begin{figure*}[t] 
   \centering
   \includegraphics[width=7in]{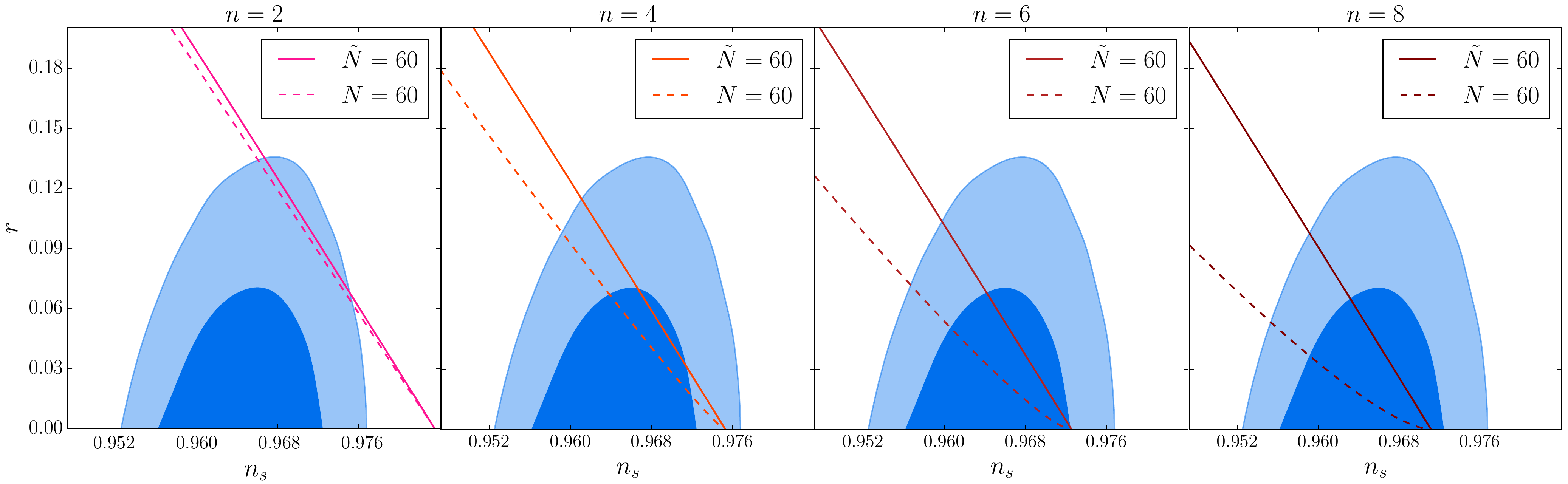} 
   \caption{Assuming $N=60$ VS $\tdN=60$ can lead to very different predictions in the $n_s$-$r$ plane. The lines show the predictions for the generalised Gaussian model $\HH=\exp(-\phi^n)$, with $n=2,4,6,8$. Solid lines show the predictions assuming $\tdN=60$. Dashed lines show those for $N=60$, which consistently underpredict $r$.}
   \label{fig_long}
\end{figure*}

Fig. \ref{fig_long} shows the comparison between the predictions assuming $\tdN=60$ (solid lines) versus $N=60$ (dashed lines). For $n=2$, as expected, the discrepancy is only roughly a few percent within the $2\sigma$ contour. This discrepancy grows quickly with $n$, and in principle the two lines can diverge even more drastically.

We conclude that the amplitude of $r$ could be vastly underestimated using the approximate expression for the number of e-folds. This severely affects the ability of future B-mode experiments (such as CORE \cite{core1,core2}) to justly accept or reject inflation models. Using $N$ to measure e-foldings, models such as \re{gen} could be falsely ruled out given a high $r$ measurement, whilst a strong constraint towards $r=0$ would be consistent with models that \ii{should have been} ruled out.




\subsection{The inflaton potential}

It is instructive to see how the potentials for the generalised Gaussian models look like. The flowchart \re{flowchart} provides an easy route towards obtaining the potential.

\begin{figure} 
   \centering
   \includegraphics[width=3.3in]{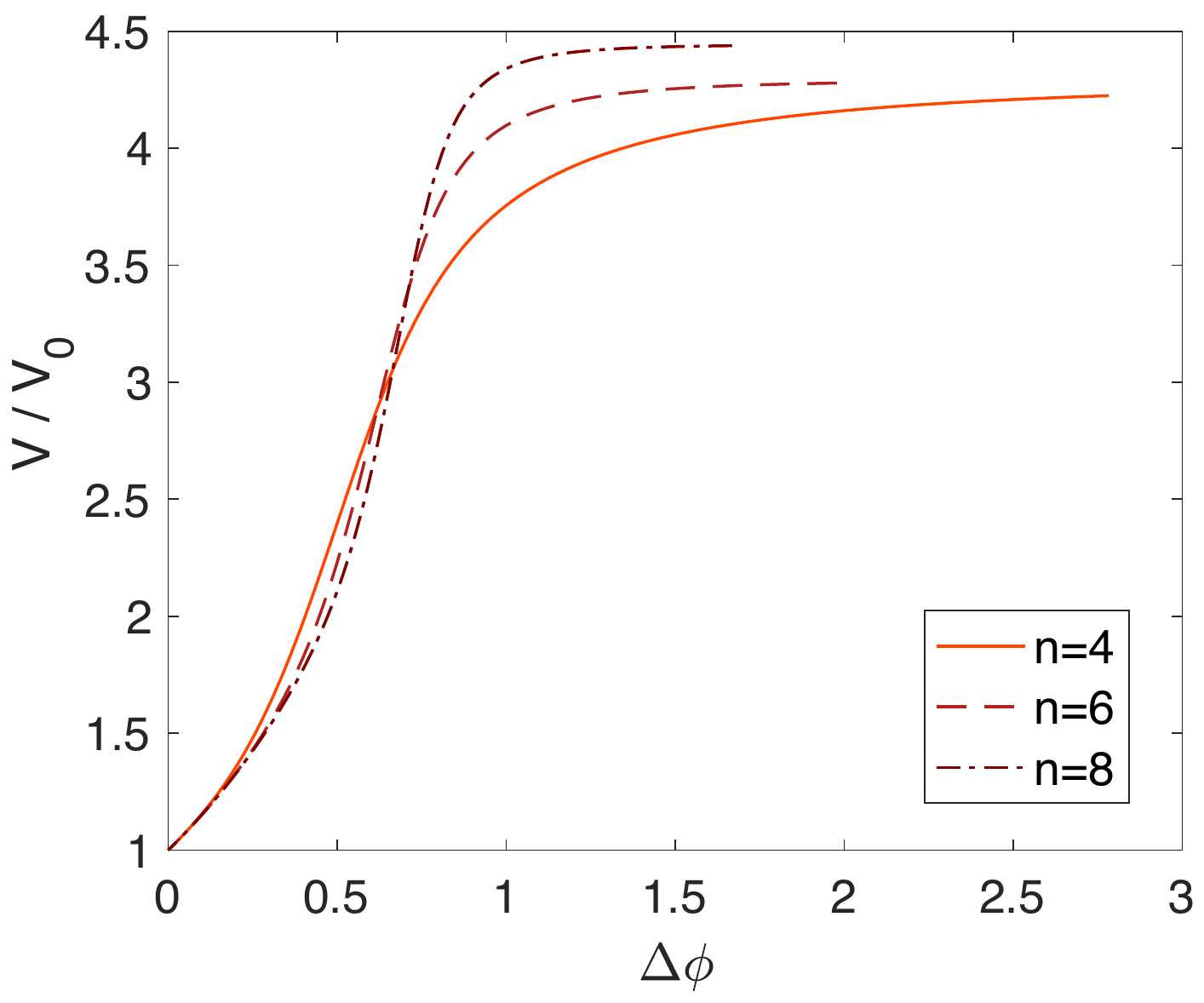} 
   \caption{The inflaton potential $V(\phi)/V(0)$ for the generalised Gaussian model $\HH=\exp(-\phi^n)$, with $n=4,6,8$. On each curve,  the bottom left corner marks the end of inflation at $\phi=0$, whilst the right-hand endpoint is where $\tdN=60$.  All these potentials are consistent with Planck's constraints in the $(n_s, r)$ plane.}
   \label{fig_potential}
\end{figure}

Fig. \ref{fig_potential} shows the potentials $V(\phi)$ for $n=4,6,8$ for a fixed value of $\alpha=1$ (we will discuss $n=2$ separately). On each curve,  inflation commences at the right-hand endpoint where $\tdN=60$, and terminating at the bottom-left corner  where $\phi\sub{end}=0$. These potentials are all consistent with Planck's $2\sigma$ constraints.

Evidently, increasing $n$ for a fixed $\alpha$ creates an increasingly steep plateau. This is consistent with our findings above that the slow-roll approximation becomes less and less accurate as $n$ increases. 

The potential for the case $n=2$ can in fact be obtained analytically. Following the flowchart, we obtain 
\ba V(\phi)&\propto (3-\beta^2)\beta^{-k}e^{k(1-\beta^2)/2},\lab{beta}\\
\text{where}\qquad \beta&=\sqrt2\alpha^2\phi+\sqrt{2\alpha^4\phi^2+1},\nn\\
k&={1\over2\alpha^2},\nn\ea
valid for $\phi\leq0$. Note that $V(\phi)$ is not an even function. In the Appendix, we prove that 
\ba V(\phi)\approx V_0 \phi^k.\ea
This proof strengthens the result in \cite{meinf} in which we showed that the predictions in the $n_s$-$r$ plane for the Gaussian $\HH(\phi)$ coincide what those of the power-law potentials. We note that for $n>2$, the potentials shown in Fig. \ref{fig_potential} can be expressed in terms of the hypergeometric function.


Had we trusted the slow-roll approximation (by setting $\eps={1\over2}(V^\pr/V)^2$), we would have been led to the slow-roll potentials $V_{SR}$:
\ba 
V_{SR}(\phi)\propto
\begin{cases}
 \phi^{1\over 2\alpha^2}, & \text{if }\ff n=2\\
 \exp\bkt{-1\over n(n-2)\alpha^n\phi^{n-2}}, & \text{if }\ff n\notin\{0, 2\}
 \end{cases}
 \ea
which may be sufficiently accurate for $n=2$, but certainly not for $n>2$. Indeed, we see that $V_{SR}$ is undefined at $\phi=0$ where inflation is supposed to end. 



\section{Gravitational-wave amplitude}

The previous section shows that using $\tdN$ or $N$ as the temporal parameter can lead to very different inflationary dynamics. In addition to inaccuracies in the predicted amplitude of $r$, we now show that the approximate e-folds can lead to an order-of-magnitude error in the predicted amplitude of primordial GW measured at laser-interferometer frequencies.

With the celebrated detections of GW from binary systems by LIGO \cite{ligo}, the hunt for GW is now progressing at a pace more fervid than ever. The most tantalising goal for the next generation of space-based laser interferometers such as BBO and DECIGO is the detection of a stochastic background of primordial GW, which would be a highly convincing evidence for an inflationary event in the early Universe. These space-based interferometry have been proposed to operate in the optimal frequency window of around 0.1$-$10 Hz, in contrast with LIGO which focuses on frequencies around $100$ Hz. See \cite{megw, buonanno, guzzetti} for reviews of direct detection of primordial GW. 

In \cite{megw}, it was shown that using the Hamilton-Jacobi formalism, the amplitude of GW from inflation can be quantified by the dimensionless energy density:
\ba 
\Omega\sub{gw}(k)h^2&\approx 4.36\times10^{-15}\,r\mc{J}(k),\\
\text{where}\quad\mc{J}(k)&\equiv\exp\bkt{-2\int_{N}^{60}\eps(N)\D N}.\notag
\ea
The lower limit, $N$, in the integral refers to the approximate e-fold number when the mode with wavenumber $k$ (or frequency $f=2\pi/k$)  exited the Hubble radius. Assume that the CMB pivot scale ($k_0=0.002$ Mpc${^{-1}}$) exited the Hubble radius at $N=60$, it follows that smaller-scale modes exited the Hubble radius at
\ba N(k)\approx 60-\ln\bkts{k\over k_0}.\ea

In contrast, using the $\HH(\tdN)$ formalism, we find that the formula for $\Omega\sub{gw}$ is now modified by replacing the damping factor $\mc{J}$ with
\ba
\tilde{\mc{J}}(k)&\equiv\exp\bkt{\int_{\tdN}^{60}\sqrt{1+2/E_1^2}\D \tdN - (60-\tdN)}.
\ea

\begin{figure*}[t] 
   \centering
   \includegraphics[width=7in]{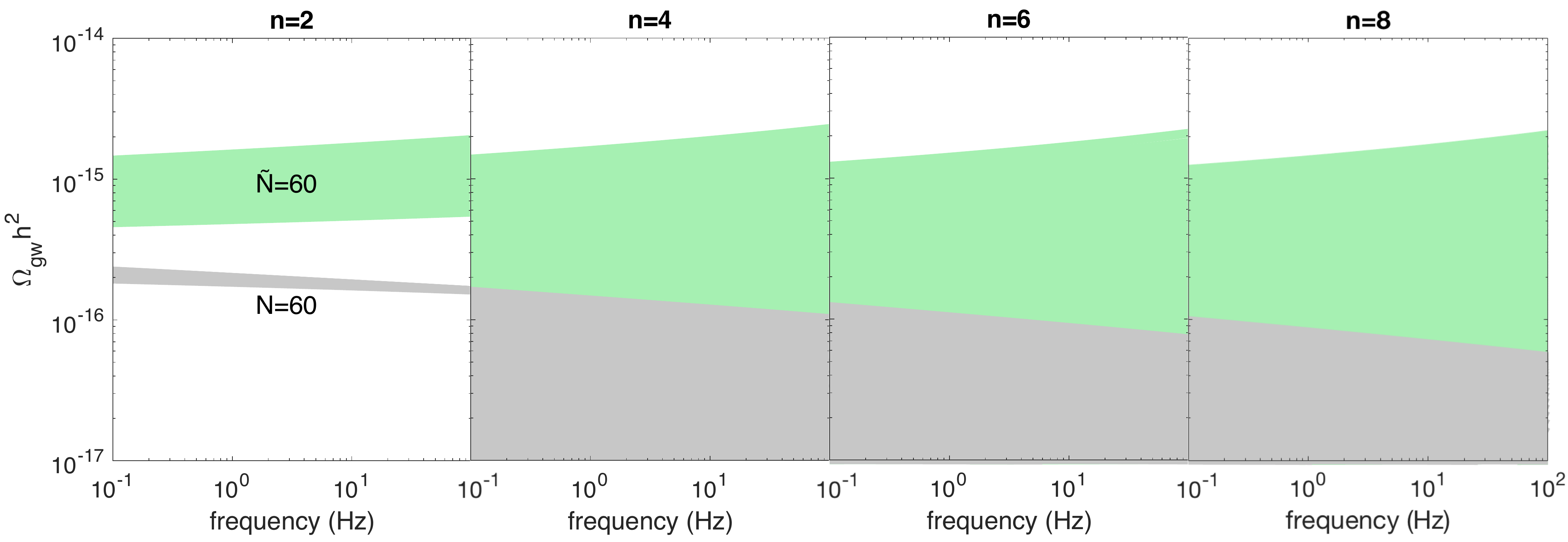} 
   \caption{Assuming $\tdN=60$ (upper/green bands) VS $N=60$ (lower/grey bands) can also lead to very different predictions for the amplitude of primordial gravitational waves, $\Omega\sub{gw}h^2$, in the frequency range $0.1-100$ Hz, where laser interferometers typically operate. These predictions are for the model \re{gen} with $n=2,4,6,8$. Except for the case $n=2$, the upper bands in fact extend downwards, intersecting with the lower grey bands.}
   \label{fig_gw}
\end{figure*}

We now compare the primordial GW predictions for the generalised Gaussian model by imposing either $\tdN=60$ or $N=60$. Fig. \ref{fig_gw} shows $\Omega\sub{gw}h^2$ in the frequency range 0.1$-$100 Hz using the values of $\alpha$ that are consistent with Planck's $2\sigma$ constraints (see Table \ref{tabalpha}). We note that for the case $n=2$, the $\tdN$ and $N$ predictions separate into clearly distinct bands, whilst for $n=4, 6, 8$, both bands extend towards 0 (\iee the two bands overlap in the grey trapezium at the bottom of each panel).

We note that the magnitudes of $\Omega\sub{gw}h^2$ spanned by the upper $\tdN=60$ bands are comparable with the putative threshold sensitivities of BBO and DECIGO (although they are sadly beyond LISA's sensitivity). Therefore, Fig \ref{fig_gw} implies that if GW were to be detected at these amplitudes, the generalised Gaussian models would be \ii{falsely ruled out} using the approximate e-folding. In other words, an order-of-magnitude underestimate in $\Omega\sub{gw}h^2$ can lead us to be overly pessimistic about the prospects for direct detection of primordial GW.



\section{Conclusion}

The future direction of inflation will be heavily influenced by the outcome of future GW experiments: Would the faint hum of primordial GW be heard at last? or would such a prospect slip further away into the next decades? Whatever the outcome may be, we must first be absolutely certain about what  we will learn from such audacious, high-risk, and hugely expensive experiments.  

The main message of this work is that using the correct measure for inflationary e-folding is crucial in deciding whether or not an inflation model could be confidently ruled out by GW experiments. As demonstrated in our detailed study of the generalised Gaussian model, using $N$ to measure e-folding is potentially very misleading, yet completely avoidable from a theoretical point of view. This suggests the widespread practice of using $N$ to study inflationary dynamics needs to be reevaluated.

Of course, there are other sources of uncertainties in the number of e-folds (the exact number in fact depends on modelling the entire history of the Universe \cite{LL}). However, using the correct measure of   e-folding is a simple and effective step to mitigate uncertainties that could affect what we will learn from future GW experiments.


\bibliographystyle{apsrev4-1}
\bibliography{inflation2}

\begin{thebibliography}{22}%
\makeatletter
\providecommand \@ifxundefined [1]{%
 \@ifx{#1\undefined}
}%
\providecommand \@ifnum [1]{%
 \ifnum #1\expandafter \@firstoftwo
 \else \expandafter \@secondoftwo
 \fi
}%
\providecommand \@ifx [1]{%
 \ifx #1\expandafter \@firstoftwo
 \else \expandafter \@secondoftwo
 \fi
}%
\providecommand \natexlab [1]{#1}%
\providecommand \enquote  [1]{``#1''}%
\providecommand \bibnamefont  [1]{#1}%
\providecommand \bibfnamefont [1]{#1}%
\providecommand \citenamefont [1]{#1}%
\providecommand \href@noop [0]{\@secondoftwo}%
\providecommand \href [0]{\begingroup \@sanitize@url \@href}%
\providecommand \@href[1]{\@@startlink{#1}\@@href}%
\providecommand \@@href[1]{\endgroup#1\@@endlink}%
\providecommand \@sanitize@url [0]{\catcode `\\12\catcode `\$12\catcode
  `\&12\catcode `\#12\catcode `\^12\catcode `\_12\catcode `\%12\relax}%
\providecommand \@@startlink[1]{}%
\providecommand \@@endlink[0]{}%
\providecommand \url  [0]{\begingroup\@sanitize@url \@url }%
\providecommand \@url [1]{\endgroup\@href {#1}{\urlprefix }}%
\providecommand \urlprefix  [0]{URL }%
\providecommand \Eprint [0]{\href }%
\providecommand \doibase [0]{http://dx.doi.org/}%
\providecommand \selectlanguage [0]{\@gobble}%
\providecommand \bibinfo  [0]{\@secondoftwo}%
\providecommand \bibfield  [0]{\@secondoftwo}%
\providecommand \translation [1]{[#1]}%
\providecommand \BibitemOpen [0]{}%
\providecommand \bibitemStop [0]{}%
\providecommand \bibitemNoStop [0]{.\EOS\space}%
\providecommand \EOS [0]{\spacefactor3000\relax}%
\providecommand \BibitemShut  [1]{\csname bibitem#1\endcsname}%
\let\auto@bib@innerbib\@empty
\bibitem [{\citenamefont {Guth}(1981)}]{guth}%
  \BibitemOpen
  \bibfield  {author} {\bibinfo {author} {\bibfnamefont {A.~H.}\ \bibnamefont
  {Guth}},\ }\href@noop {} {\bibfield  {journal} {\bibinfo  {journal} {Phys.
  Rev.}\ }\textbf {\bibinfo {volume} {D23}},\ \bibinfo {pages} {347} (\bibinfo
  {year} {1981})}\BibitemShut {NoStop}%
\bibitem [{\citenamefont {Linde}(1982)}]{linde}%
  \BibitemOpen
  \bibfield  {author} {\bibinfo {author} {\bibfnamefont {A.}~\bibnamefont
  {Linde}},\ }\href {\doibase http://dx.doi.org/10.1016/0370-2693(82)91219-9}
  {\bibfield  {journal} {\bibinfo  {journal} {Physics Letters B}\ }\textbf
  {\bibinfo {volume} {108}},\ \bibinfo {pages} {389 } (\bibinfo {year}
  {1982})}\BibitemShut {NoStop}%
\bibitem [{\citenamefont {Starobinsky}(1982)}]{starobinsky2}%
  \BibitemOpen
  \bibfield  {author} {\bibinfo {author} {\bibfnamefont {A.~A.}\ \bibnamefont
  {Starobinsky}},\ }\href@noop {} {\bibfield  {journal} {\bibinfo  {journal}
  {Phys. Lett. B}\ }\textbf {\bibinfo {volume} {117}},\ \bibinfo {pages} {175}
  (\bibinfo {year} {1982})}\BibitemShut {NoStop}%
\bibitem [{\citenamefont {Liddle}\ and\ \citenamefont {Lyth}(2000)}]{LLbook}%
  \BibitemOpen
  \bibfield  {author} {\bibinfo {author} {\bibfnamefont {A.~R.}\ \bibnamefont
  {Liddle}}\ and\ \bibinfo {author} {\bibfnamefont {D.~H.}\ \bibnamefont
  {Lyth}},\ }\href {\doibase 10.1017/CBO9781139175180} {\emph {\bibinfo {title}
  {Cosmological Inflation and Large-Scale Structure}}}\ (\bibinfo  {publisher}
  {Cambridge University Press},\ \bibinfo {year} {2000})\BibitemShut {NoStop}%
\bibitem [{\citenamefont {{Baumann}}(2009)}]{baumann}%
  \BibitemOpen
  \bibfield  {author} {\bibinfo {author} {\bibfnamefont {D.}~\bibnamefont
  {{Baumann}}},\ }\href@noop {} {\bibfield  {journal} {\bibinfo  {journal}
  {ArXiv e-prints}\ } (\bibinfo {year} {2009})},\ \Eprint
  {http://arxiv.org/abs/0907.5424} {arXiv:0907.5424} \BibitemShut {NoStop}%
\bibitem [{\citenamefont {{Martin}}\ \emph {et~al.}(2014)\citenamefont
  {{Martin}}, \citenamefont {{Ringeval}},\ and\ \citenamefont
  {{Vennin}}}]{martin}%
  \BibitemOpen
  \bibfield  {author} {\bibinfo {author} {\bibfnamefont {J.}~\bibnamefont
  {{Martin}}}, \bibinfo {author} {\bibfnamefont {C.}~\bibnamefont
  {{Ringeval}}}, \ and\ \bibinfo {author} {\bibfnamefont {V.}~\bibnamefont
  {{Vennin}}},\ }\href {\doibase 10.1016/j.dark.2014.01.003} {\bibfield
  {journal} {\bibinfo  {journal} {Physics of the Dark Universe}\ }\textbf
  {\bibinfo {volume} {5}},\ \bibinfo {pages} {75} (\bibinfo {year}
  {2014})}\BibitemShut {NoStop}%
\bibitem [{Note1()}]{Note1}%
  \BibitemOpen
  \bibinfo {note} {The amplitude of superhorizon perturbations is frozen at
  least in the standard models of inflation presented in this paper. See \cite
  {leach} for interesting case studies.}\BibitemShut {Stop}%
\bibitem [{\citenamefont {{Chongchitnan}}(2016)}]{meinf}%
  \BibitemOpen
  \bibfield  {author} {\bibinfo {author} {\bibfnamefont {S.}~\bibnamefont
  {{Chongchitnan}}},\ }\href {\doibase 10.1103/PhysRevD.94.043526} {\bibfield
  {journal} {\bibinfo  {journal} {\prd}\ }\textbf {\bibinfo {volume} {94}},\
  \bibinfo {eid} {043526} (\bibinfo {year} {2016})}\BibitemShut {NoStop}%
\bibitem [{lis()}]{lisa}%
  \BibitemOpen
  \href@noop {} {}\bibinfo {howpublished}
  {\url{https://www.elisascience.org}}\BibitemShut {NoStop}%
\bibitem [{\citenamefont {{Corbin}}\ and\ \citenamefont
  {{Cornish}}(2006)}]{bbo}%
  \BibitemOpen
  \bibfield  {author} {\bibinfo {author} {\bibfnamefont {V.}~\bibnamefont
  {{Corbin}}}\ and\ \bibinfo {author} {\bibfnamefont {N.~J.}\ \bibnamefont
  {{Cornish}}},\ }\href {\doibase 10.1088/0264-9381/23/7/014} {\bibfield
  {journal} {\bibinfo  {journal} {Classical and Quantum Gravity}\ }\textbf
  {\bibinfo {volume} {23}},\ \bibinfo {pages} {2435} (\bibinfo {year}
  {2006})}\BibitemShut {NoStop}%
\bibitem [{dec()}]{decigo}%
  \BibitemOpen
  \href@noop {} {}\bibinfo {howpublished}
  {{\url{http://tamago.mtk.nao.ac.jp/decigo/index\_E.html}}}\BibitemShut
  {NoStop}%
\bibitem [{\citenamefont {{Planck Collaboration}}(2016)}]{planck15}%
  \BibitemOpen
  \bibfield  {author} {\bibinfo {author} {\bibnamefont {{Planck
  Collaboration}}},\ }\href {\doibase 10.1051/0004-6361/201525830} {\bibfield
  {journal} {\bibinfo  {journal} {A\&A}\ }\textbf {\bibinfo {volume} {594}},\
  \bibinfo {eid} {A13} (\bibinfo {year} {2016})}\BibitemShut {NoStop}%
\bibitem [{\citenamefont {Liddle}\ \emph {et~al.}(1994)\citenamefont {Liddle},
  \citenamefont {Parsons},\ and\ \citenamefont {Barrow}}]{liddle+}%
  \BibitemOpen
  \bibfield  {author} {\bibinfo {author} {\bibfnamefont {A.~R.}\ \bibnamefont
  {Liddle}}, \bibinfo {author} {\bibfnamefont {P.}~\bibnamefont {Parsons}}, \
  and\ \bibinfo {author} {\bibfnamefont {J.~D.}\ \bibnamefont {Barrow}},\
  }\href@noop {} {\bibfield  {journal} {\bibinfo  {journal} {Phys. Rev.}\
  }\textbf {\bibinfo {volume} {D50}},\ \bibinfo {pages} {7222} (\bibinfo {year}
  {1994})}\BibitemShut {NoStop}%
\bibitem [{\citenamefont {Lidsey}\ \emph {et~al.}(1997)\citenamefont {Lidsey},
  \citenamefont {Liddle}, \citenamefont {Kolb}, \citenamefont {Copeland},
  \citenamefont {Barreiro},\ and\ \citenamefont {Abney}}]{lidseybig}%
  \BibitemOpen
  \bibfield  {author} {\bibinfo {author} {\bibfnamefont {J.~E.}\ \bibnamefont
  {Lidsey}}, \bibinfo {author} {\bibfnamefont {A.~R.}\ \bibnamefont {Liddle}},
  \bibinfo {author} {\bibfnamefont {E.~W.}\ \bibnamefont {Kolb}}, \bibinfo
  {author} {\bibfnamefont {E.~J.}\ \bibnamefont {Copeland}}, \bibinfo {author}
  {\bibfnamefont {T.}~\bibnamefont {Barreiro}}, \ and\ \bibinfo {author}
  {\bibfnamefont {M.}~\bibnamefont {Abney}},\ }\href@noop {} {\bibfield
  {journal} {\bibinfo  {journal} {Rev. Mod. Phys.}\ }\textbf {\bibinfo {volume}
  {69}},\ \bibinfo {pages} {373} (\bibinfo {year} {1997})}\BibitemShut
  {NoStop}%
\bibitem [{\citenamefont {{COrE Collaboration}}(2011)}]{core1}%
  \BibitemOpen
  \bibfield  {author} {\bibinfo {author} {\bibnamefont {{COrE
  Collaboration}}},\ }\href@noop {} {\bibfield  {journal} {\bibinfo  {journal}
  {ArXiv e-prints}\ } (\bibinfo {year} {2011})},\ \Eprint
  {http://arxiv.org/abs/1102.2181} {arXiv:1102.2181} \BibitemShut {NoStop}%
\bibitem [{\citenamefont {{CORE Collaboration}}(2016)}]{core2}%
  \BibitemOpen
  \bibfield  {author} {\bibinfo {author} {\bibnamefont {{CORE
  Collaboration}}},\ }\href@noop {} {\bibfield  {journal} {\bibinfo  {journal}
  {ArXiv e-prints}\ } (\bibinfo {year} {2016})},\ \Eprint
  {http://arxiv.org/abs/1612.08270} {arXiv:1612.08270} \BibitemShut {NoStop}%
\bibitem [{lig()}]{ligo}%
  \BibitemOpen
  \href@noop {} {}\bibinfo {howpublished}
  {\url{https://www.ligo.caltech.edu}}\BibitemShut {NoStop}%
\bibitem [{\citenamefont {{Chongchitnan}}\ and\ \citenamefont
  {{Efstathiou}}(2006)}]{megw}%
  \BibitemOpen
  \bibfield  {author} {\bibinfo {author} {\bibfnamefont {S.}~\bibnamefont
  {{Chongchitnan}}}\ and\ \bibinfo {author} {\bibfnamefont {G.}~\bibnamefont
  {{Efstathiou}}},\ }\href {\doibase 10.1103/PhysRevD.73.083511} {\bibfield
  {journal} {\bibinfo  {journal} {\prd}\ }\textbf {\bibinfo {volume} {73}},\
  \bibinfo {eid} {083511} (\bibinfo {year} {2006})}\BibitemShut {NoStop}%
\bibitem [{\citenamefont {{Buonanno}}\ and\ \citenamefont
  {{Sathyaprakash}}(2014)}]{buonanno}%
  \BibitemOpen
  \bibfield  {author} {\bibinfo {author} {\bibfnamefont {A.}~\bibnamefont
  {{Buonanno}}}\ and\ \bibinfo {author} {\bibfnamefont {B.~S.}\ \bibnamefont
  {{Sathyaprakash}}},\ }\href@noop {} {\bibfield  {journal} {\bibinfo
  {journal} {ArXiv e-prints}\ } (\bibinfo {year} {2014})},\ \Eprint
  {http://arxiv.org/abs/1410.7832} {arXiv:1410.7832} \BibitemShut {NoStop}%
\bibitem [{\citenamefont {{Chiara Guzzetti}}\ \emph {et~al.}(2016)\citenamefont
  {{Chiara Guzzetti}}, \citenamefont {{Bartolo}}, \citenamefont {{Liguori}},\
  and\ \citenamefont {{Matarrese}}}]{guzzetti}%
  \BibitemOpen
  \bibfield  {author} {\bibinfo {author} {\bibfnamefont {M.}~\bibnamefont
  {{Chiara Guzzetti}}}, \bibinfo {author} {\bibfnamefont {N.}~\bibnamefont
  {{Bartolo}}}, \bibinfo {author} {\bibfnamefont {M.}~\bibnamefont
  {{Liguori}}}, \ and\ \bibinfo {author} {\bibfnamefont {S.}~\bibnamefont
  {{Matarrese}}},\ }\href@noop {} {\bibfield  {journal} {\bibinfo  {journal}
  {ArXiv e-prints}\ } (\bibinfo {year} {2016})},\ \Eprint
  {http://arxiv.org/abs/1605.01615} {arXiv:1605.01615} \BibitemShut {NoStop}%
\bibitem [{\citenamefont {Liddle}\ and\ \citenamefont {Leach}(2003)}]{LL}%
  \BibitemOpen
  \bibfield  {author} {\bibinfo {author} {\bibfnamefont {A.~R.}\ \bibnamefont
  {Liddle}}\ and\ \bibinfo {author} {\bibfnamefont {S.~M.}\ \bibnamefont
  {Leach}},\ }\href@noop {} {\bibfield  {journal} {\bibinfo  {journal} {Phys.
  Rev.}\ }\textbf {\bibinfo {volume} {D68}},\ \bibinfo {pages} {103503}
  (\bibinfo {year} {2003})}\BibitemShut {NoStop}%
\bibitem [{\citenamefont {{Leach}}\ and\ \citenamefont
  {{Liddle}}(2001)}]{leach}%
  \BibitemOpen
  \bibfield  {author} {\bibinfo {author} {\bibfnamefont {S.~M.}\ \bibnamefont
  {{Leach}}}\ and\ \bibinfo {author} {\bibfnamefont {A.~R.}\ \bibnamefont
  {{Liddle}}},\ }\href {\doibase 10.1103/PhysRevD.63.043508} {\bibfield
  {journal} {\bibinfo  {journal} {\prd}\ }\textbf {\bibinfo {volume} {63}},\
  \bibinfo {eid} {043508} (\bibinfo {year} {2001})}\BibitemShut {NoStop}%
\end{thebibliography}%

\section*{Appendix -- Proof that the Gaussian model corresponds to power-law inflation}

Let $n=2$ in Eq. \ref{gen}. We want to show that $V\propto\phi^k$. Let $x\equiv\sqrt2 a^2\phi$. Thus the goal is equivalent to showing that 
$$\diff{\ln V}{\ln x}=k.$$
From the definition of $\beta=x+\sqrt{x^2+1}$ (Eq. \ref{beta}), the inverse function is
$$x={\beta-\beta^{-1}\over2}.$$
Let $\lambda=\ln\beta$, so that $x=\sinh\lambda$. Note that since $\phi<0$, $\beta$ is a small positive number, and so $\lambda$ is a large negative number. 

Taking the log of Eq. \re{beta} and differentiating wrt $\lambda$ gives
$$\diff{\ln V}{\lambda}\approx -k,$$
where we ignore the small contributions from the terms $\sim e^{2\lambda}$. Therefore,
$$\diff{\ln V}{\ln x}=\diff{\ln V}{\lambda}\diff{\lambda}{\ln x}\approx-k\tanh\lambda\approx k,$$
since $\tanh\lambda\approx -1$. Hence the power-law correspondence is proved.

\end{document}